\documentclass[12pt]{article}
\usepackage{epsf,amsmath, amsfonts}
\usepackage{epsfig,mathtools}
\usepackage{color,graphicx}
\usepackage{dcolumn}
\usepackage{bm}
\usepackage{epsfig,latexsym}
\usepackage{lineno,newtxtext,nicefrac}
\usepackage{dashrule,hyperref}
\usepackage{mathptmx}      

\usepackage[skip=0pt]{caption}
\usepackage[skip=0pt]{subcaption}
\newcommand{\bseq}{\begin{subequations}}
\newcommand{\eseq}{\end{subequations}}
\newcommand{\beq}{\begin{equation}}
\newcommand{\eeq}{\end{equation}}
\newcommand{\bef}{\begin{figure}}
\newcommand{\eef}{\end{figure}}

\hypersetup{
    colorlinks=true,
    urlcolor=red,
}

\begin{document}

\title{\large \bf Spatio temporal linear stability of viscoelastic free shear flows: I (dilute regime)} 
\author{ {\bf S. Sircar$\dagger, \ddagger$} and {\bf D. Bansal$\dagger$}
\\{$\dagger$ \small Department of Mathematics, IIIT Delhi, India 110020} \\{\small $\ddagger$ Corresponding Author (email: sarthok@iiitd.ac.in)}}
\maketitle

\begin{abstract}
\noindent We report the temporal and spatio-temporal stability analyses of anti-symmetric, free shear, viscoelastic flows obeying the Oldroyd-B constitutive equation in the limit of low to moderate Reynolds number and Weissenberg number. The resulting fourth order Orr-Sommerfeld equation is reduced to a set of six auxiliary equations which are numerically integrated starting from the rescaled far-field conditions, i.~e., via the Compound Matrix Method. Numerical results indicate that with increasing Weissenberg number: (a) the peak of the maximal growth rate (i. e., the maximum value of the imaginary component of the growth rate) is reduced, (b) the entire range of unstable spectrum is shifted towards longer waves (i.~e., the entire region of temporal instability is gradually concentrated near zero wavenumber), (c) the vorticity structure contours are dilated and (d) the residual Reynolds stresses are diminished. All these observations suggest that viscoelasticity reduces the temporal instability but does not completely suppress it. The Briggs idea of analytic continuation is deployed to classify regions of temporal stability, absolute and convective instabilties as well as evanescent (false) modes, in a finite range of Reynolds number, Weissenberg number and the viscosity coefficient. The main result is that, unlike Newtonian fluids, the free shear flow of dilute polymeric liquids are either (absolutely/convectively) unstable for all Reynolds number or the transition to instability occurs at very low Reynolds number, a finding attributed to the fact that viscoelasticity aggravates free surface flow instabilities. Although this transitional pathway connecting the temporally stable state to the elastoinertially unstable state have been identified by some in-vitro experiments, but until now, has not been quantified theoretically via linear stability analysis.

\end{abstract}


\section{Introduction} \label{sec:intro}
Understanding the hydrodynamic stability and transition of free shear flows of dilute polymer solutions is paramount to both the fundamental theory of viscoelastic liquids~\cite{Bird1987Vol1,Bird1987Vol2} and their industrial applications, especially those arising in microfluidic mixing~\cite{Squires2005}, viscoelastic thin films~\cite{Tomar2006} and viscoelastic stabilization via polymer addition~\cite{White2008}. Likewise, understanding the mechanism governing the transition to turbulence in shearing flows of viscoelastic biofluids is decisive in resolving the swelling instability transition in mucus~\cite{Sircar2016DCDS}, cartilage~\cite{Sircar2015}, adhesion-fragmentation transition in cells~\cite{Sircar2013,Sircar2014,Sircar2016JOMB} and bending instability transition in soft tissues~\cite{Destrade2009}. Classical approaches to the viscoelastic instability studies involve identifying an equilibrium state, whose stability is studied through eigenvalue analysis by linearizing the governing equations. The eigenvalue analysis seek the least stable eigenmode of the linearized mass and momentum conservation equations suitably transformed to the Orr-Sommerfeld Equation (OSE)~\cite{Drazin2004,TKS2016} via parallel flow approximation. One of the features of traditional eigenvalue analysis is that the disturbance field is assumed to grow either in space or in time. Huerre and Monkewitz~\cite{Huerre1990} have attempted to apply the combined spatio-temporal theory to a restricted class of Newtonian mixing layers with the goal of determining a general criterion whereby a class of flows can be analyzed by either the spatial or temporal theory by inspecting the dispersion relation (DRP) in the complex wave number-frequency plane. Some other applications as well as in-vivo and in-silico stability studies of shear flows of Newtonian and non-Newtonian liquids are listed in~\cite{Batchelor2002,Schmid2012,Kalliadasis2012}. However, the models and experiments listed in these references fail to present a comprehensive, spatio-temporal analyses of plane, free shear flows in the form of an instability-phase diagram. Therefore a motivation of this article is to illustrate this phase diagram in the viscoelastic parameter space.

The characterization of fluid flows as being absolutely or convectively unstable and the method of spatio-temporal analysis by progressive moving of the isocontours in the complex frequency and wavenumber plane was first proposed by Briggs~\cite{Briggs1964} and later by Bers~\cite{Bers1983} in the context of plasma physics. Rallison {\it et al.} detailed separate studies of temporal as well as spatial instability in channel flows of dilute suspensions, including stratified flows~\cite{Wilson1998} and multilayer Couette and Poiseuille flows~\cite{Miller2007}, obeying either the Oldroyd-B, Upper Convected Maxwell or the FENE fluid constitutive equations. The literature on theoretical studies of the spatio-temporal linear analysis of viscoelastic flows are more recent. Govindarajan highlighted the role of the spatio-temporal hierarchical structure of complex fluids, in determining the alterations of flow stability~\cite{Govindarajan2014}. A series of early~\cite{Hansen1974} and later experiments in viscoelastic, wall bounded flows~\cite{Poole2016}, extensional flows~\cite{Varshney2016}, torsional flows~\cite{Haward2016} and in Taylor-Couette flows~\cite{Schafer2013}, outline the transition to turbulence at Reynolds numbers ($Re$) much lower than the Newtonian threshold, thereby dubbing this phenomena as `early turbulence'. Consequently, this `new' transitional pathway, connecting the laminar to a novel elastoinertial turbulent state is speculated to be either captured only by full Direct Numerical Simulations (DNS)~\cite{Samanta2013} or by purely elastic subcritical (or non-linear) instability analysis~\cite{Pan2013}. However, this article offers a crucial conceptual explanation of the above mentioned transitional pathway to early turbulence via a linear, spatio-temporal stability analysis.

Through linear stability analysis, we aim to address the following intriguing questions: What is the critical flow / polymer relaxation conditions for the onset of instability ? and more crucially What is the linear spatio-temporal, time asymptotic response of the flow at the critical value of the material parameters ? In the next section, we delineate the model of the free shear layer flow coupled with the Oldroyd-B constitutive relation for the extra elastic stress tensor (\S \ref{subsec:modelA}) and the details of the linear stability analysis via the fourth order OSE (\S \ref{subsec:modelB}). A thorough description of the Briggs contour integral method to determine the existence of absolute instability, convective instability and evanescent modes in spatio-temporal analysis in included in this section. In \S \ref{sec:CMM}, we introduce the Compound Matrix Method to numerically integrate the resultant system of stiff differential equations for the auxiliary variables emerging from the Orr-Sommerfeld equation. Section 4 highlights the simulation results of the temporal (\S \ref{subsec:time}) and the spatio-temporal (\S \ref{subsec:space-time}) stability analyses of anti-symmetric, free shear flows of polymeric liquids. This article concludes with a brief discussion of the implication of these results and the focus of our future direction (\S \ref{sec:conclusion}).

\section{Mathematical model and linear stability analysis} \label{sec:model}
Unlike Newtonian solvents, the stability predictions of polymeric liquids depend upon the details of the equations relating stress to the shear rate. The Oldroyd-B constitutive equations has its range of applicability limited to dilute solutions and moderate shear rates~\cite{Bird1987Vol2}. It predicts no shear thinning, a constant first normal stress coefficient and a zero second normal stress coefficient. The numerical solution obtained by linearizing the Navier Stokes along with the extra stress constitutive equation (or the OSE stability equation) is reliable in describing the characteristics of the initial stages of the mixing layer transition~\cite{Bird1987Vol1}. The linear stability results can be utilized as the initial-boundary conditions for DNS~\cite{TKS2016} or the large eddy simulation of polymeric liquids~\cite{Samanta2013} in order to further study the flow evolving process downstream.

\subsection{Problem definition} \label{subsec:modelA}
For the mixing layer flow configuration it is customary to assume $\mathcal{U}_1$ (respectively $\mathcal{U}_2$) as the free-stream velocity of the upper (lower) flow. We denote $\mathcal{U}_0 = \frac{1}{2}(\mathcal{U}_1-\mathcal{U}_2)$ as the free-stream velocity in a reference frame moving with the average velocity of the flow (i. e., $\frac{1}{2}(\mathcal{U}_1+\mathcal{U}_2)$) and $\delta$, the momentum thickness of the mixing layer. The continuity and the momentum equations for an incompressible flow are,
\beq
\nabla\cdot {\bf v} = 0, \qquad \rho \frac{D {\bf v}}{D t} = -\nabla p + {\bf v} \cdot \tau,  
\label{eqn:MainModel} 
\eeq
where ${\bf v}$ is the velocity vector, $\rho$ the density, $p$ the isotropic pressure and $\tau$ the extra stress tensor. Equations \eqref{eqn:MainModel} are closed through the evolution equation for the extra stress tensor $\tau$, written as the sum of the viscous, Newtonian stress, $\tau^s$ (= $\eta_s {\bf D}$, ${\bf D}$ is the shear rate tensor) and the elastic stress, $\tau^p$ (= $\eta_p {\bf A}$), where $\eta_s, \eta_p$ are the solvent viscosity and the polymeric contribution to the shear viscosity, respectively. We let $\eta (= \eta_s + \eta_p)$ and $\nu (= \eta_s/\eta)$ denote the total viscosity and the viscous contribution to the total viscosity of the fluid, respectively. The constitutive equation for the extra stress tensor, $\tau$, obeying the Oldroyd-B model then becomes,
\beq
\tau = \eta [\nu {\bf D} + (1-\nu){\bf A}], 
\label{eqn:Stress}
\eeq
where the tensor ${\bf A}$ satisfies the Upper Convected Maxwell equation,
\beq
\frac{\partial A}{\partial t} + {\bf v} \cdot \nabla {\bf A} - \nabla {\bf v}^T \cdot {\bf A} - {\bf A} \cdot \nabla {\bf v} = \frac{{\bf D} - {\bf A}}{\lambda},
\label{eqn:Oldroyd}
\eeq
$\lambda$ is the polymer relaxation time. 

\subsection{Linear Stability Analysis} \label{subsec:modelB}
Using $\mathcal{U}_0$ and $\delta$ as the reference velocity and length scale, respectively, we characterize the non-dimensional form of equations~\eqref{eqn:MainModel}, by the dimensionless numbers, $Re = \nicefrac{\rho \delta \mathcal{U}_0}{\eta}$, and the Weissenberg number, $We = \nicefrac{\lambda \mathcal{U}_0}{\delta}$. We assume that the mean flow is two-dimensional and quasi-parallel with its variation entirely in the direction normal to the flow, i.~e., 
\beq
U(y) = \tanh(y), \,\, \Omega(y) = \tanh^2(y)-1, \,\, \Psi(y) = \log(\cosh(y)),
\label{eqn:Meanflow}
\eeq
where $U(y), \Omega(y), \Psi(y)$ are the dimensionless streamwise mean velocity (with zero spanwise mean velocity), spanwise mean vorticity and the associated streamfunction, respectively. Further, assume that the mean flow supports a two-dimensional disturbance field. The streamfunction and the extra stress tensor are represented by the base state profile $(\Psi(y), \mathbf{T}(y))$ plus a small perturbation, Fourier transformed in $x$ and $t$ as follows,
\beq
\psi(x,y,t) = \Psi(y) + \phi(y) e^{{\it i}(\alpha x - \omega t)}, \quad \tau(x,y,t) = \mathbf{T}(y) + \varphi(y) e^{{\it i}(\alpha x - \omega t)}, 
\label{eqn:disturbance}
\eeq
where $\phi(y), \varphi(y)$ are the spanwise perturbations in the streamfunction and the extra stress tensor and $\alpha, \omega$ are the complex wavenumber and angular frequency, respectively. Rewriting equation~\eqref{eqn:MainModel} in the streamfunction-vorticity formulation and utilizing equations~\eqref{eqn:Stress}, \eqref{eqn:Oldroyd}, \eqref{eqn:Meanflow}, \eqref{eqn:disturbance}, we arrive at the equation governing the perturbation of the streamfunction, given by the familiar fourth order OSE~\cite{Azaiez1994},
\bseq \label{eqn:OSE}
\begin{align}
i\big\{(\alpha U-\omega)(\phi''-\alpha^{2}\phi)-\alpha U'' \phi \big\}-\frac{\nu}{Re}\left((\frac{d}{dy})^{2}-\alpha^{2})\right)^{2}\phi = \frac{1-\nu}{\mathcal{F}Re}\sum_{n=0}^{4} c_{n} \phi^{(n)},
\tag{\ref{eqn:OSE}}
\end{align} 
where $( )'$ and $( )^{(n)}$ denote the first and the n$^{\text th}$ derivative of $\phi$ with respect to $y$ and the coefficients $c_i$'s are
\begin{align}
\begin{split}
c_{0}(y) & = \alpha^{4} + \mathcal{F}^{(4)}-\alpha^{2}\bigg(\frac{\mathcal{F}''}{\mathcal{F}}\bigg)(\mathcal{F}-1) - 2\alpha^{2}\bigg(\frac{\mathcal{F}'}{\mathcal{F}}\bigg)^{2}(\mathcal{F}^{2}+1) - 4\bigg(\frac{\mathcal{F}'}{\mathcal{F}}\bigg)\mathcal{F}''' - 3\frac{(\mathcal{F}'')^{2}}{\mathcal{F}}\\
& +~ 4\frac{(\mathcal{F}')^{4}}{\mathcal{F}^{2}} - 6 \mathcal{F}'' \bigg(\frac{\mathcal{F}'}{\mathcal{F}}\bigg)^{2}(\mathcal{F} - 1),\\
c_{1}(y) & = -2\alpha^{2}\bigg(\frac{\mathcal{F}'}{\mathcal{F}}\bigg)(\mathcal{F}-1) + 4\bigg(\frac{\mathcal{F}''}{\mathcal{F}}\bigg)\bigg(\frac{\mathcal{F}'}{\mathcal{F}}\bigg)(\mathcal{F}-1)^{2} - 4\bigg(\frac{\mathcal{F}'}{\mathcal{F}}\bigg)^{2}(\text{D}_y\mathcal{F})(\mathcal{F} - 1)\\
& +~ 2\bigg(\frac{\mathcal{F}'''}{\mathcal{F}}\bigg)(\mathcal{F} - 1),\\
c_{2}(y) & = -2\alpha^{2} + 3\bigg(\frac{\mathcal{F}''}{\mathcal{F}}\bigg)(\mathcal{F} - 1) + 2\bigg(\frac{\mathcal{F}'}{\mathcal{F}}\bigg)^{2}(\mathcal{F} - 1)^{2},\\
c_{3}(y) & = 2\bigg(\frac{\mathcal{F}'}{\mathcal{F}}\bigg)(\mathcal{F}-1),\\
c_{4}(y) & = 1,
\end{split}
\label{eqn:coefficients}
\end{align}
\eseq
and $\mathcal{F} = 1 + i We(\alpha U-\omega)$, solved together with the vanishing boundary conditions for the free shear flow disturbance ($\phi \rightarrow 0, \phi' \rightarrow 0$ as $y \rightarrow \pm \infty$). Limiting the domain of integration to the upper half of the flow and analyzing only the anti-symmetric disturbance, the boundary conditions are revised as follows,
\bseq
\begin{align}
\phi = \phi'' = 0 \quad &\text{at} \quad y = 0 \label{eqn:zeroBC}\\
\phi = \phi' = 0 \quad &\text{at} \quad y \rightarrow \infty. \label{eqn:farfieldBC}
\end{align}
\eseq
In \S~\ref{sec:CMM}, we delineate a numerical solution procedure for solving the eigenvalue equation~\eqref{eqn:OSE} together with the boundary conditions~\eqref{eqn:zeroBC}\eqref{eqn:farfieldBC}. In the rest of this section we describe a mechanism to classify the spatio-temporal flow instabilities.


The spatio-temporal evolution of a localized disturbance (located at the origin of the $x$-$t$ plane) is illustrated by considering the response of a given base velocity profile, $u(x, t)$, to an impulse excitation~\cite{Huerre1990},
\beq
D\left( -{\it i} \frac{\partial }{\partial x}, {\it i} \frac{\partial }{\partial t}, {\bf M} \right) u(x, t) = \delta(x) \delta(t), 
\label{eqn:DRP}
\eeq
where $D(\alpha, \omega, {\bf M})=0$ is the DRP relation and ${\bf M}$ is the vector of material and fluid parameters. The solution to equation~\eqref{eqn:DRP} is dictated by the Green's function, 
\beq
G(x, t) = {\displaystyle \frac{1}{4\pi^2} \int_L \int_F \frac{e^{{\it i}(\alpha x - \omega t)}}{D(\alpha, \omega, {\bf M})} d\alpha d\omega }
\label{eqn:Green}
\eeq
where $F$ and $L$ are the Fourier contour in $\alpha$ and the Laplace contour in the $\omega$ plane. The Fourier contour integral is placed parallel to the Real($\alpha$) axis (in the ensuing description we denote real/imaginary components with subscript {\it r}/{\it i}, respectively), while the Laplace contour integral is placed above all singularities of the DRP in the $\omega$-plane so as to satisfy the causality condition (i. e., $G \equiv 0$ if $t < 0$). An important criteria for the understanding of instability entails the study of the flow behavior in the `long term' (i. e., $t \rightarrow \infty$) where an analytical solution of equation~\eqref{eqn:DRP} is possible. In the asymptotic limit of long time, the integration of equation~\eqref{eqn:Green} is analytically accomplished by using the method of stationary phase~\cite{Ablowitz2003}, i.~e.,
\beq
G(x, t) \sim -\frac{1}{\sqrt{2\pi}} \frac{e^{{\it i}[\pi/4 + \alpha_* x - \omega_* t]}}{\frac{\partial D}{\partial \omega}\left[ \frac{d^2\omega}{d \alpha^2} \right]^{1/2}},
\label{eqn:ApproxGreen}
\eeq
where $\alpha_*$ is the saddle point in the $\alpha$ plane (i.~e., the root of $\frac{\partial D}{\partial \alpha}=0$) and $\omega_*$ is the corresponding branch point in the $\omega$ plane satisfying the DRP relation. From equation~\eqref{eqn:ApproxGreen}, we can surmise the condition for which the flow will be {\it absolutely unstable} (signifying the growth of disturbance in both upstream and downstream direction from the origin, otherwise known as the `resonance mode'~\cite{Lingwood1997}), i.~e.,
\beq
G(x ,t) \xrightarrow[]{t \rightarrow \infty} \infty, \qquad \text{along the ray} \,\, {\displaystyle \nicefrac{x}{t}} = 0,
\eeq
versus when the flow will be {\it convectively unstable} (where disturbances are swept downstream from the source and given sufficient time these disturbances decay at any fixed position in space, also known as the `driven mode'), i.~e.,
\beq
G(x ,t) \xrightarrow[]{t \rightarrow \infty} 0, \qquad \text{along the ray} \,\, \nicefrac{x}{t} = 0.
\eeq
Our analysis also reveal the presence of evanescent modes (or false modes) in the flow field, which is the non-propagating mode or the locally concentrated mode~\cite{Drazin1979}, described later in this section.

The necessary (but not sufficient) condition for the presence of absolute instability is the vanishing characteristic of the group velocity, ${\it v}_g$, at the saddle point in the $\alpha$-plane or the branch point in the $\omega$-plane (${\it v}_g = \frac{\partial \omega}{\partial \alpha} = \nicefrac{(\frac{\partial D}{\partial \alpha})}{(\frac{\partial D}{\partial \omega})} = 0$ such that $\omega=D(\alpha)$). But the group velocity is zero at every saddle point, especially where the two $\alpha$-branches meet, independent of whether the branches originate from the same half of the $\alpha$-plane (i. e., when evanescent modes are detected) or not. To overcome this inadequacy, Briggs~\cite{Briggs1964} devised the idea of analytic continuation in which the Laplace contour $L$, in equation~\eqref{eqn:Green}, is deformed towards the $\omega_r$ axis of the complex $\omega$-plane, with the simultaneous adjustment of the Fourier contour $F$ in the $\alpha$-plane to maintain the separation of the $\alpha$-branches; those which originate from the top half (the upstream modes with $\alpha_i > 0$) from those which originate from the bottom half of the $\alpha$-plane (or the downstream modes). The deformation of the $F$ contour (while preserving causality) is inhibited, however, when the paths of the two $\alpha$-branches originating from the opposite halves of the $\alpha$-plane intersect each other, leading to the appearance of saddle points which are the {\it pinch point}, $\alpha^{\text pinch}$. The concurrent branch point appearance in the $\omega$-plane is the {\it cusp point}, $\omega^{\text cusp}$ (i. e., $D(\alpha^{\text pinch}, \omega^{\text cusp})\!=\!\frac{\partial D(\alpha^{\text pinch}, \omega^{\text cusp})}{\partial \alpha}\!=\!0$ but $\frac{\partial^2 D(\alpha^{\text pinch}, \omega^{\text cusp})}{\partial \alpha^2}\!\ne \!0$). Kupfer~\cite{Kupfer1987} employed this local mapping procedure to conceptualize the stability characteristics of this branch point. Near a `reasonably close' neighborhood of the pinch point, a local Taylor expansion yields a DRP relation which has a second-order algebraic form in the $\omega$-plane (and which is a first order saddle point in the $\alpha$-plane), i.~e., $(\omega - \omega^{\text cusp}) \sim (\alpha - \alpha^{\text pinch})^2$. This period-doubling characteristics of the map causes the $\alpha_i$-contours to `rotate' around $\omega^{\text cusp}$, forming a cusp. In the $\omega$-plane, we draw a ray parallel to the $\omega_i$-axis from the cusp point such that it intersects the image of the F-contour (or $\alpha_i = 0$ curve) and count the number of intersections (consequently, count the number of times both $\alpha$-branches cross the $\alpha_r$-axis before forming a pinch point in the $\alpha$-plane, as shown in Figure~\ref{fig:Figure5}). If the ray drawn from the cusp point intersects the image of the F contour in the $\omega$-plane (or if either one or both the $\alpha$-branches cross the $\alpha_r$-axis) even number of times, then the flow dynamics correspond to an evanescent mode. Otherwise, in the case of odd intersections the observed cusp point is genuine, leading to either absolutely unstable system (in the upper half of the $\omega$-plane) or convectively unstable system (in the lower half of the $\omega$-plane); provided the system is temporally unstable.

\section{Solution to the eigenvalue problem} \label{sec:CMM}
The eigenvalues (i.~e., the values of ($\alpha, \omega$) satisfying equation~\eqref{eqn:OSE}) are found by examining the consequence of the far stream boundary condition, given by equation~\eqref{eqn:farfieldBC}, on the solution structure of the OSE. In the limit $y \rightarrow \infty$, we have $U(y)=1$ and $U''(y)=0$, which reduces the OSE to the following constant coefficient ODE~\cite{Drazin2004},
\beq
\phi^{(4)} - 2\alpha^2 \phi'' + \alpha^4 \phi = \frac{{\it i} Re \mathcal{F}_\infty}{1-\nu + \nu \mathcal{F}_\infty} (\alpha - \omega)(\phi'' - \alpha^2 \phi),
\label{eqn:OSE-CC}
\eeq
where $\mathcal{F}_\infty=1+We(\alpha - \omega)$. The solution to equation~\eqref{eqn:OSE-CC} can be obtained in the form $\phi = e^{\lambda y}$, with the characteristic roots given by $\lambda_{1,2}=\mp \alpha$ and $\lambda_{3,4}=\mp q$, where $q = {\displaystyle \left[ \alpha^2 + \frac{{\it i} Re \mathcal{F}_\infty}{1-\nu + \nu \mathcal{F}_\infty} (\alpha - \omega)\right]^{1/2}}$. The fourth order OSE~\eqref{eqn:OSE} will have four fundamental solutions, $\{ \phi_i \}^4_{i=1}$, i.~e., $\phi = {\displaystyle \sum^4_{i=1} a_i \phi_i }$. To satisfy the boundary condition~\eqref{eqn:farfieldBC}, one must have $a_2=a_4=0$ for real ($\alpha, q$) > 0 which warrants a general solution of the form
\beq
\phi = a_1 \phi_1 + a_3 \phi_3.
\label{eqn:general}
\eeq
Equation~\eqref{eqn:general} is a non-trivial, admissible solution of the OSE, satisfying the zero boundary conditions at the centerline (i. e., at $y=0$, equation~\eqref{eqn:zeroBC}) if and only if the determinant of the associated matrix of the linear algebraic system vanishes at $y=0$, i. e., 
\beq
\left(\phi_1 \phi''_3 - \phi''_1 \phi_3\right)|_{y=0} = 0,
\label{eqn:OSE_DRP}
\eeq
which is the dispersion relation of the problem. In the asymptotic limit of $Re \rightarrow \infty$, the eigenmodes of the free shear layer instability problems are such that $|q| \gg |\alpha|$. This enormous contrast between the two sets of roots of the characteristic equation~\eqref{eqn:OSE-CC} is the source of stiffness causing the fundamental solutions of the OSE to vary by several orders of magnitude in the entire physical domain, and thus necessitate the use of Compound Matrix Method (CMM)~\cite{Ng1985}. In CMM, one works with a set of auxiliary variables which are combinations of the fundamental solutions $\phi_1$ and $\phi_3$, namely
\begin{align}
y_1 &= \phi_1 \phi'_3 - \phi_3 \phi'_1, \nonumber \\
y_2 &= \phi_1 \phi''_3 - \phi_3 \phi''_1, \nonumber \\
y_3 &= \phi_1 \phi'''_3 - \phi_3 \phi'''_1, \nonumber \\
y_4 &= \phi'_1 \phi''_3 - \phi''_1 \phi'_3, \nonumber \\
y_5 &= \phi'_1 \phi'''_3 - \phi'''_1 \phi'_3, \nonumber \\
y_6 &= \phi''_1 \phi'''_3 - \phi'''_1 \phi''_3,
\label{eqn:AuxiliaryVar}
\end{align}
satisfying the corresponding first order ODEs,
\begin{align}
y'_1 &= y_2, \nonumber\\
y'_2 &= y_3 + y_4, \nonumber\\
y'_3 &= y_5 - \frac{(1-\nu)(c_1 y_1 + c_2 y_2 + c_3 y_3) + \mathcal{F}({\it i}\text{Re}(\alpha U - \omega)+2\alpha^2\nu)y_2}{(1-\nu + \mathcal{F}\nu)}, \nonumber\\
y'_4 &= y_5, \nonumber\\
y'_5 &= y_6 + \frac{(1-\nu)(c_0 y_1 - c_2 y_4 - c_3 y_5) + \mathcal{F}({\it i}\text{Re}(\alpha U - \omega)+2\alpha^2\nu)y_4}{(1-\nu + \mathcal{F}\nu)} \nonumber\\
& + \frac{\mathcal{F}({\it i}\text{Re}[ (\alpha U - \omega)\alpha^2 + \alpha U''] + \alpha^4\nu)y_1}{(1-\nu + \mathcal{F}\nu)}, \nonumber\\
y'_6 &= \frac{(1-\nu)(c_0 y_2 + c_1 y_4 - c_3 y_6) + \mathcal{F}({\it i}\text{Re}[ (\alpha U - \omega)\alpha^2 + \alpha U''] + \alpha^4\nu)y_2}{(1-\nu + \mathcal{F}\nu)}. 
\label{eqn:Auxilliary}
\end{align}
Since, at $y \rightarrow \infty$: $\phi_1 \sim e^{-\alpha y}$ and $\phi_3 \sim e^{-q y}$, we can estimate the free stream values of the unknowns,
\begin{align}
y_1 &\sim (-q+\alpha) e^{-(\alpha+q)y}, \nonumber \\
y_2 &\sim (q^2-\alpha^2) e^{-(\alpha+q)y}, \nonumber \\
y_3 &\sim (-q^3+\alpha^3) e^{-(\alpha+q)y}, \nonumber \\
y_4 &\sim (-\alpha q^2+\alpha^2 q) e^{-(\alpha+q)y}, \nonumber \\
y_5 &\sim (\alpha q^3-\alpha^3 q) e^{-(\alpha+q)y}, \nonumber \\
y_6 &\sim (-\alpha^2 q^3+\alpha^3 q^2) e^{-(\alpha+q)y}.
\label{eqn:FarField}
\end{align}
Since all the variables in equation~\eqref{eqn:FarField} attenuate at the same rate, the stiffness of the problem is removed by rescaling the variables, say with respect to $y_1$, so that the (rescaled) free-stream conditions for solving equations~\eqref{eqn:Auxilliary} are
\begin{align}
y_1 &= 1.0, \nonumber \\
y_2 &= -(\alpha +q), \nonumber \\
y_3 &= \alpha^2 +q\alpha + q^2, \nonumber \\
y_4 &= q\alpha, \nonumber \\
y_5 &= -q\alpha(\alpha + q), \nonumber \\
y_6 &= (q\alpha)^2
\label{eqn:ICs}
\end{align}
The solution for equations~\eqref{eqn:Auxilliary} is obtained by marching backward from the free stream to the centerline, with the initial conditions provided by equations~\eqref{eqn:ICs}. A suitable value of ($\alpha, \omega$), or the eigenvalue for a given combination of material parameters {\bf M} is obtained by enforcing the DRP relation (equation~\eqref{eqn:OSE_DRP}) in auxiliary variables, i.~e.,
\beq
y_2 = 0 \quad \text{at} \quad y = 0. \label{eqn:DRPAux}
\eeq
The eigenfunction, $\phi$, are found using a suitable linear combination of $\phi_1$ and $\phi_3$, via equations~\eqref{eqn:general}, \eqref{eqn:AuxiliaryVar}, \eqref{eqn:Auxilliary}, such that the characteristic roots at free stream (i. e., the solution to equation~\eqref{eqn:OSE-CC}) have a negative real part and the coefficients in the ODE do not vanish at the centerline. An equation satisfying these constraints was suggested by Davey~\cite{Davey1980} as follows,
\beq
y_4 \phi''' - y_5 \phi'' + y_6 \phi' = 0.
\label{eqn:eigenFns}
\eeq
%
\section{Results} \label{sec:result}
In our {\it in-silico} experiments, the zeros of the DRP relation (equation~\eqref{eqn:OSE_DRP}, \eqref{eqn:DRPAux}) were found in the $\alpha / \omega$ plane inside the region $|\omega_r| \le 1.5, |\omega_i| \le 1.5, |\alpha_r| \le 3.0$ and $|\alpha_i| \le 0.1$, using a discrete step-size of $\Delta \alpha_{\text r/i} = \Delta \omega_{\text r/i} = 5 \times 10^{-3}$. The free stream boundary conditions were applied at $\eta=12$, which leads to the numerical value of the momentum thickness, $\delta=0.30685$. Numerical integration from the free stream to the shear layer centerline was accomplished via the fourth order Runge Kutta time integration with discrete step-size of $\Delta y = 6.67 \times 10^{-4}$ in the tranverse direction. The eigenfunctions are found by numerically integrating equation~\eqref{eqn:eigenFns} from the centerline to the free stream starting from the centerline conditions, equation~\eqref{eqn:zeroBC}. The presence of only two centerline conditions imply that the eigenfunctions are scaled with respect to the centerline value of $\phi'$ (i .e., $\phi'(0)$). Hence, the results discussed in \S \ref{subsec:time}, \ref{subsec:space-time}, involving these eigenfunctions, are qualitative.

\subsection{Temporal stability analysis} \label{subsec:time}
\begin{figure}[htbp]
\centering
\includegraphics[scale=0.15]{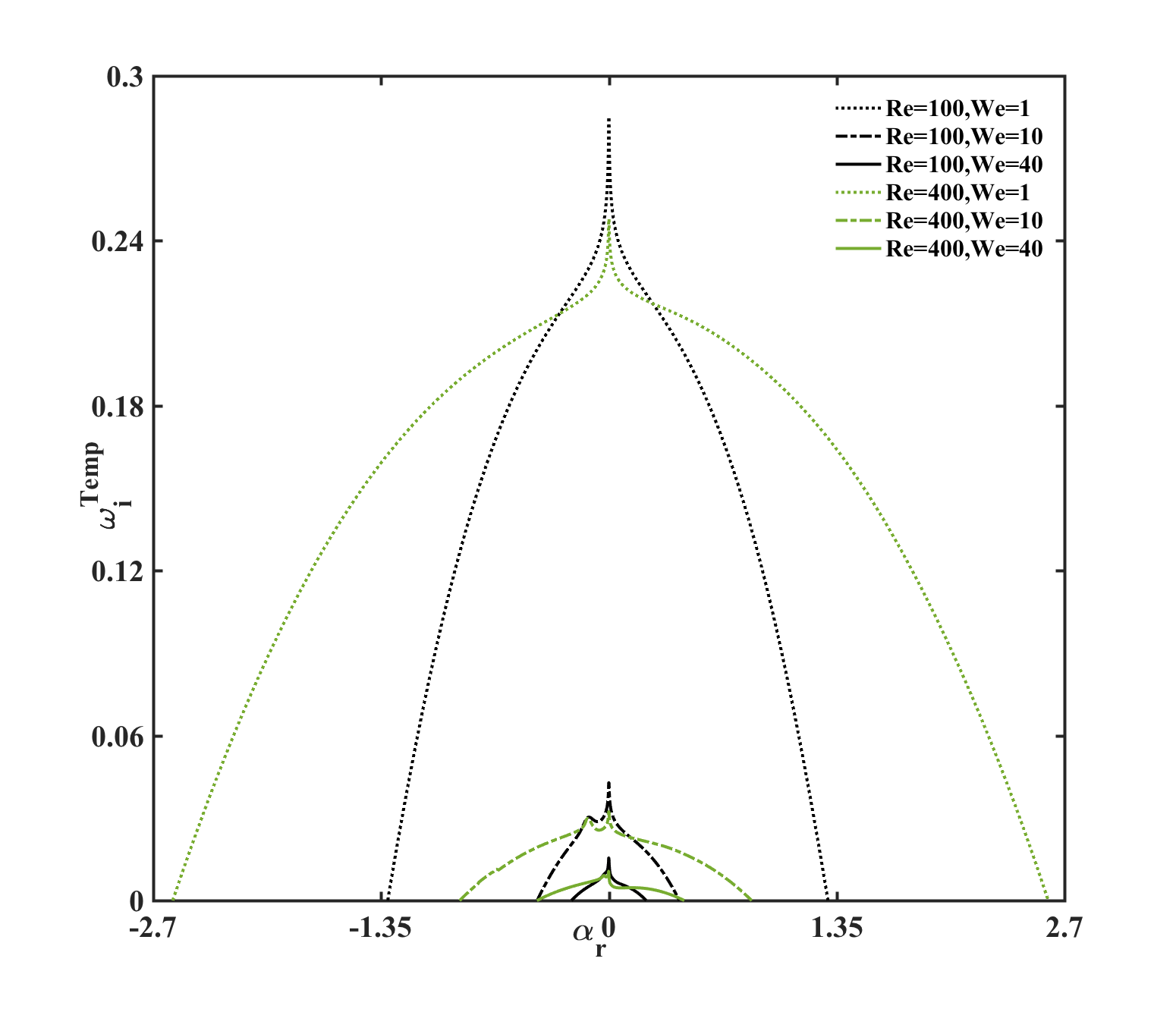}
  \vspace*{\fill}%
\caption{Maximum temporal growth rate of the disturbance, $\omega^{\text{Temp}}_i$ versus wavenumber, $\alpha_r$ for Weissenberg number, $We=1$ (dashed curve), $We=10$ (dash-dotted curve), $We=40$ (dotted curve) and for Reynolds number, $Re=100$ (black curve), $Re=400$ (green curve). The peak of these curves are at $\alpha^{\text{max}}=-0.001$ and $\omega^{\text{max}}=$0.784+0.285{\it i} (for $Re=100, We=1$); 0.084+0.043{\it i} (for $Re=100, We=10$); 0.015+0.016{\it i} (for $Re=100, We=40$); 0.795+0.248{\it i} (for $Re=400, We=1$); 0.070+0.032{\it i} (for $Re=400, We=10$); 0.020+0.011{\it i} (for $Re=400, We=40$). The viscosity coefficient is fixed at $\nu=0.5$. }\label{fig:Fig1}
\end{figure}

In this section we explore the stability of the solution of the OSE (equation~\eqref{eqn:OSE}) by exclusively assigning $\omega$ to be a complex number. The temporal stability analysis in the inviscid limit was earlier studied by Azaiez~\cite{Azaiez1994}, but owing to the limitations in their analysis some of the conclusions obtained in their work need to be revisited. In Figure~\ref{fig:Fig1} we present the maximum temporal growth rate of the disturbance (i.~e., $\omega^{\text{Temp}}_i$, the largest positive imaginary component of any root of the DRP relation~\eqref{eqn:OSE_DRP}) versus the wavenumber, $\alpha_r$. Note that, as $We$ is increased, the peak of the maximum growth rate, $\omega^{\text{max}}_i$ as well as the range of the unstable wavenumbers (i.~e., the range of $\alpha_r$ in which $\omega^{\text{Temp}}_i > 0$) is reduced and the entire temporally unstable spectrum is shifted towards longer waves (i.~e., the region of instability is gradually concentrated around $\alpha_r=0$). All these observations suggest a mechanism of elastic stabilization in non-Newtonian fluids which we have discussed below. Later, in \S \ref{subsec:space-time}, we outline how the nature of this instability changes from absolute instability to convective instability. 
%

\begin{figure}[htbp]
\centering
\begin{subfigure}{.5\textwidth}
  \centering
 \hspace*{\fill}%
  \includegraphics[width=.9\linewidth]{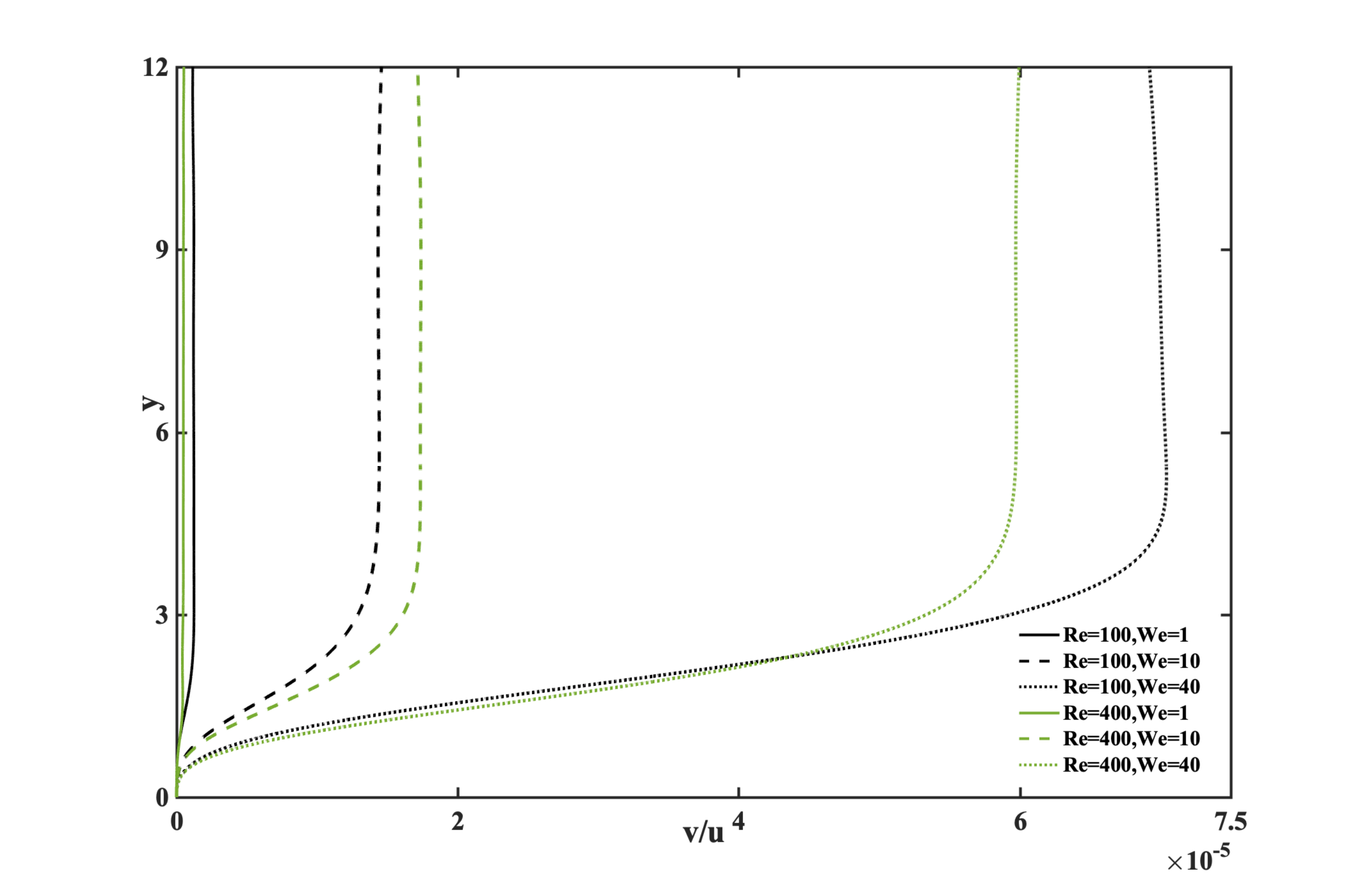}
  \subcaption{ }
  \label{fig:Fig2a}
\end{subfigure}%
\begin{subfigure}{.5\textwidth}
\vskip 15pt
  \centering
  \hspace*{\fill}%
  \includegraphics[width=.9\linewidth]{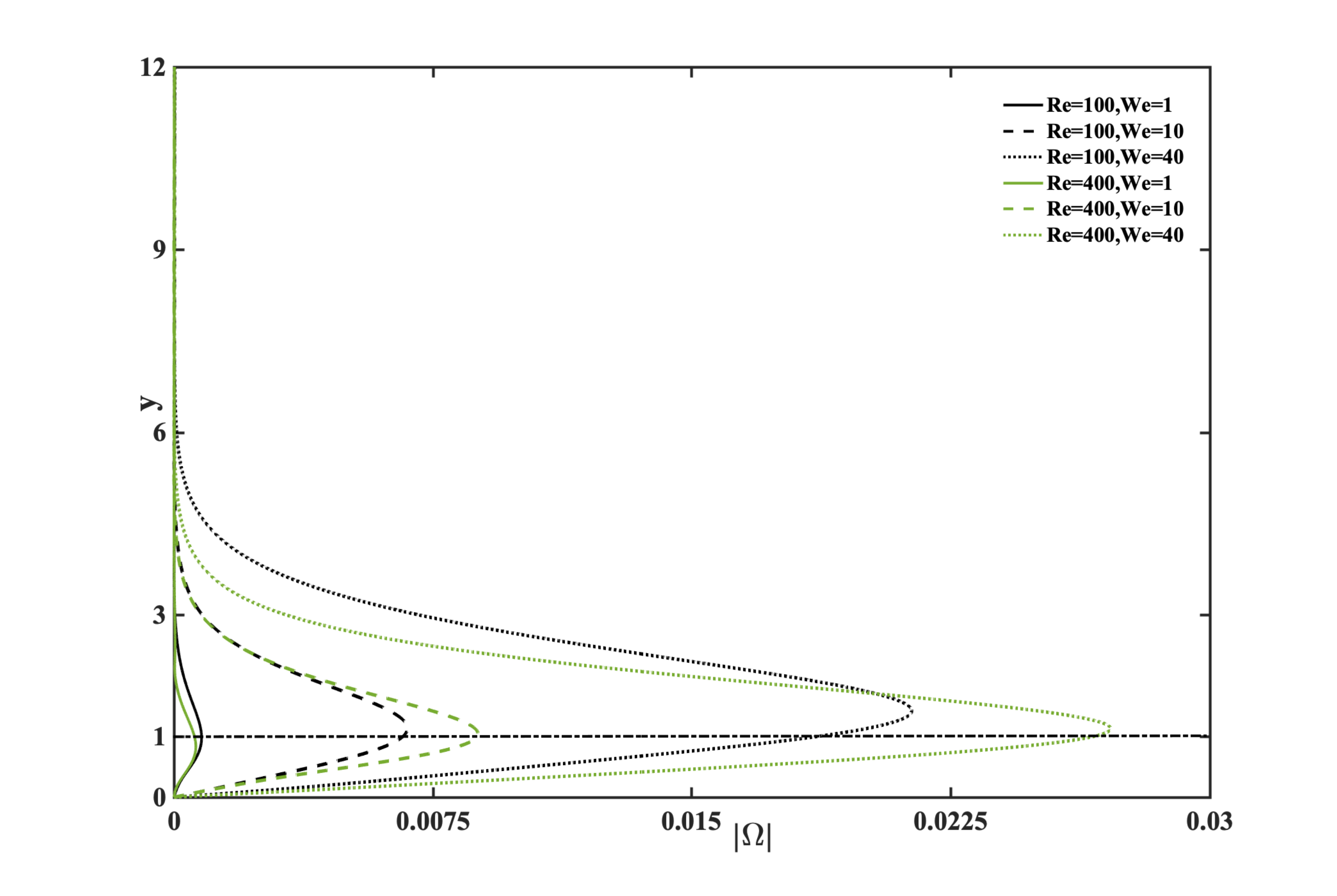}
  \subcaption{ }
  \label{fig:Fig2b}
\end{subfigure}
\caption{(a) The ratio of the transverse vs streamwise velocity disturbance, ${\displaystyle \frac{u}{v}=-\frac{\phi'}{\alpha \phi}}$ and (b) the variation of the magnitude of the vorticity disturbance, ${\displaystyle |\omega|=|\alpha^2 \phi - \phi''|}$ 
in the transverse direction (refer equation~\eqref{eqn:disturbance}), evaluated at the peak of the maximum growth rate curves ($\alpha^{\text max}, \omega^{\text max}$) whose values are listed in Figure~\ref{fig:Fig1} caption. A horizontal line is drawn to distinguish the mixing region, $y \le 1$. All curves are numerically estimated at the viscosity coefficient $\nu=0.5$.}\label{fig:Fig2}
\end{figure}

Further characterization of the presence of viscoelasticity on the flow is highlighted via the ratio of the transverse to the streamwise velocity fluctuations (Figure~\ref{fig:Fig2a}) and the variations of the magnitude of the vorticity disturbance (Figure~\ref{fig:Fig2b}). These figures are presented for six different combinations of ($Re, We$) evaluated at the peak of the maximum growth rate curves, $\alpha^{\text{max}}, \omega^{\text{max}}$  (refer Figure~\ref{fig:Fig1} caption for the respective numerical values). Figure~\ref{fig:Fig2a} depicts the relative strength of the transverse velocity fluctuations versus tranverse spatial coordinates. We remark that at a lower value of $We$ (e.~g., $We=1.0$) these relative fluctuations are significant within the viscous mixing region ($y \le 1$) whereas at higher $We$ values (e.~g., $We=40.0$) these fluctuations become important outside the mixing layer region and in a region where the fluid is essentially moving at the free-stream velocity (i.~e., at $We=40.0$ the velocity fluctuation ratio attains its maximum at $y \approx 3.8$ where the mean velocity U$_0(y)$ is 99\% of its free-stream value). This study suggests that in traditionally non-Newtonian shear flows, the dynamical effects of the disturbance on the mean flow are negligible when compared to an equivalent Newtonian case~\cite{TKS2016}. This observation is further corroborated by observing the magnitude of the vorticity disturbance (Figure~\ref{fig:Fig2b}). While at lower $We$ ($We \le 10.0$) the region of maximum magnitude of vorticity disturbance is inside the range $y \in [0, 1]$, at higher $We$ this maximum occurs in the range, $y \ge 1$.

\begin{figure}[htbp]
\centering
\begin{subfigure}{.5\textwidth}
  \centering
  \hspace*{\fill}%
  \includegraphics[width=\linewidth, height=0.7\linewidth]{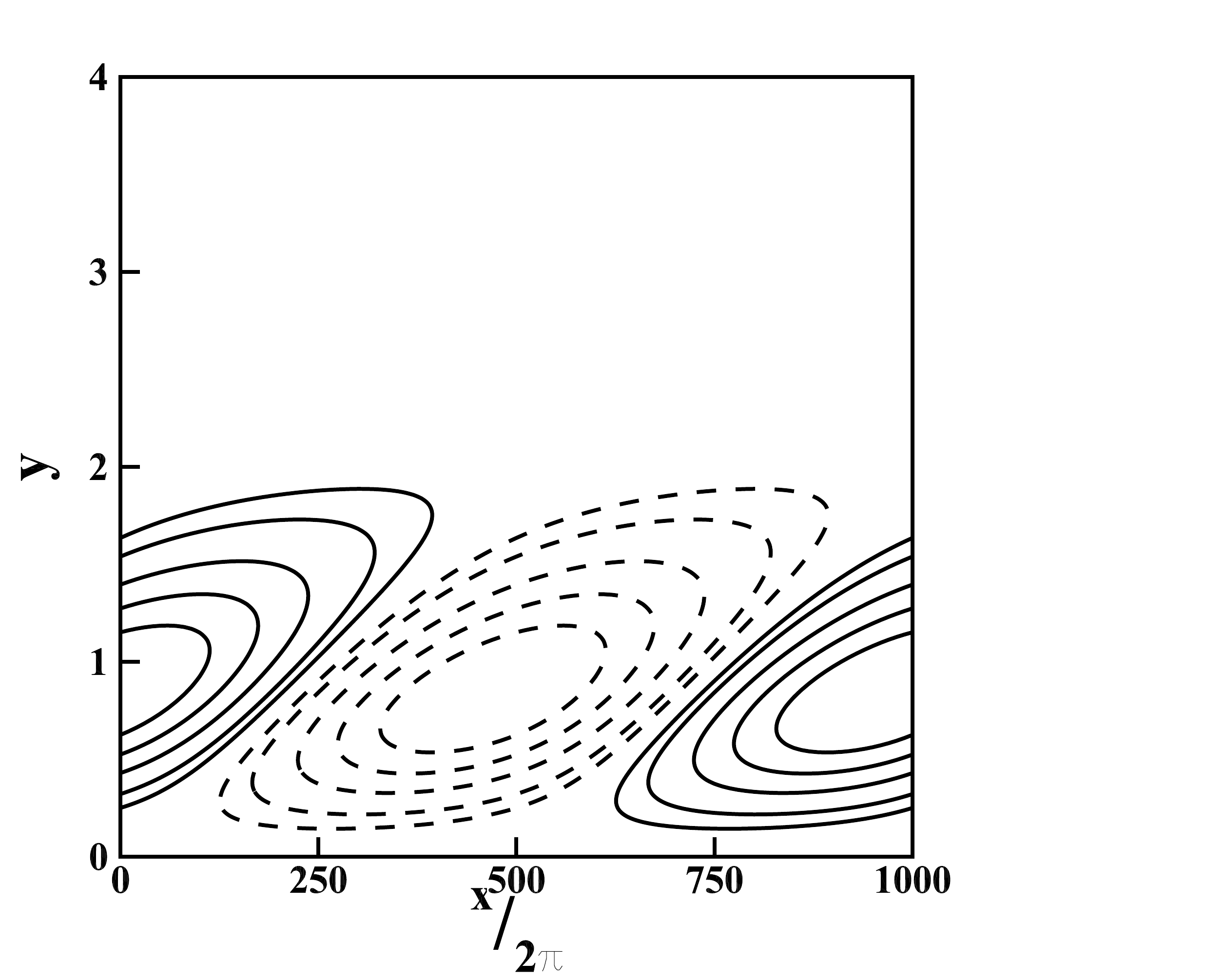}
  \subcaption{ }
  \label{fig:Fig3a}
\end{subfigure}%
\begin{subfigure}{.5\textwidth}
  \centering
  \hspace*{\fill}%
  \includegraphics[width=\linewidth, height=0.7\linewidth]{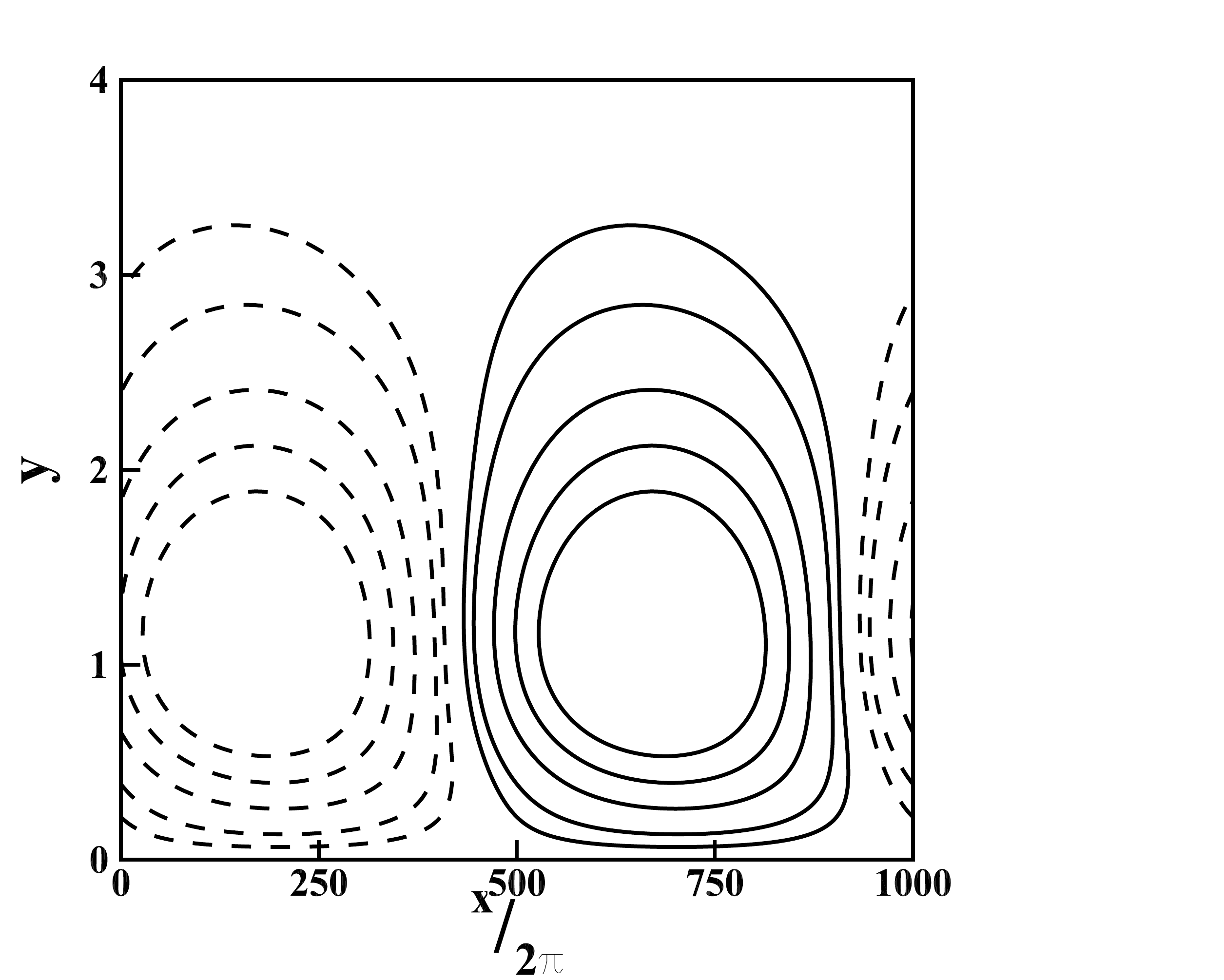}
  \subcaption{ }
  \label{fig:Fig3b}
\end{subfigure}
\caption{The rate of vorticity production contours ${\displaystyle \frac{D\Omega}{Dt}(x,y) = {\it i}(D_{yy}\phi-\alpha^2\phi)(\omega-U\alpha)e^{{\it i}\alpha x}}$ in the streamwise versus transverse direction at (a) $Re=400, We=1$, and (b) $Re=400, We=40$, evaluated at the peak of the maximum growth rate curves (refer Figure~\ref{fig:Fig1} caption). The contours are drawn at identically equal levels and the viscosity coefficient is fixed at $\nu=0.5$ in both cases.}\label{fig:Figure3}
\end{figure}

More insights about this mechanism of stabilization can be gained by comparing the contours of the rate of vorticity production (drawn at identically equal levels) for an arbitrary time in the streamwise versus transverse direction, at $Re=400, We=1.0$ (Figure~\ref{fig:Fig3a}) and $Re=400, We=40.0$ (Figure~\ref{fig:Fig3b}) evaluated at the eigenvalues corresponding to the peak of the growth rate curves (refer Figure~\ref{fig:Fig1} caption). Note that as $We$ increases, the vorticity {\it structures} become larger and the contour spacing becomes wider, thereby suggesting an enhanced drag reduction. 

\begin{figure}[htbp]
\begin{subfigure}{.5\textwidth}
  \centering
  \includegraphics[width=.9\linewidth, height=0.75\linewidth]{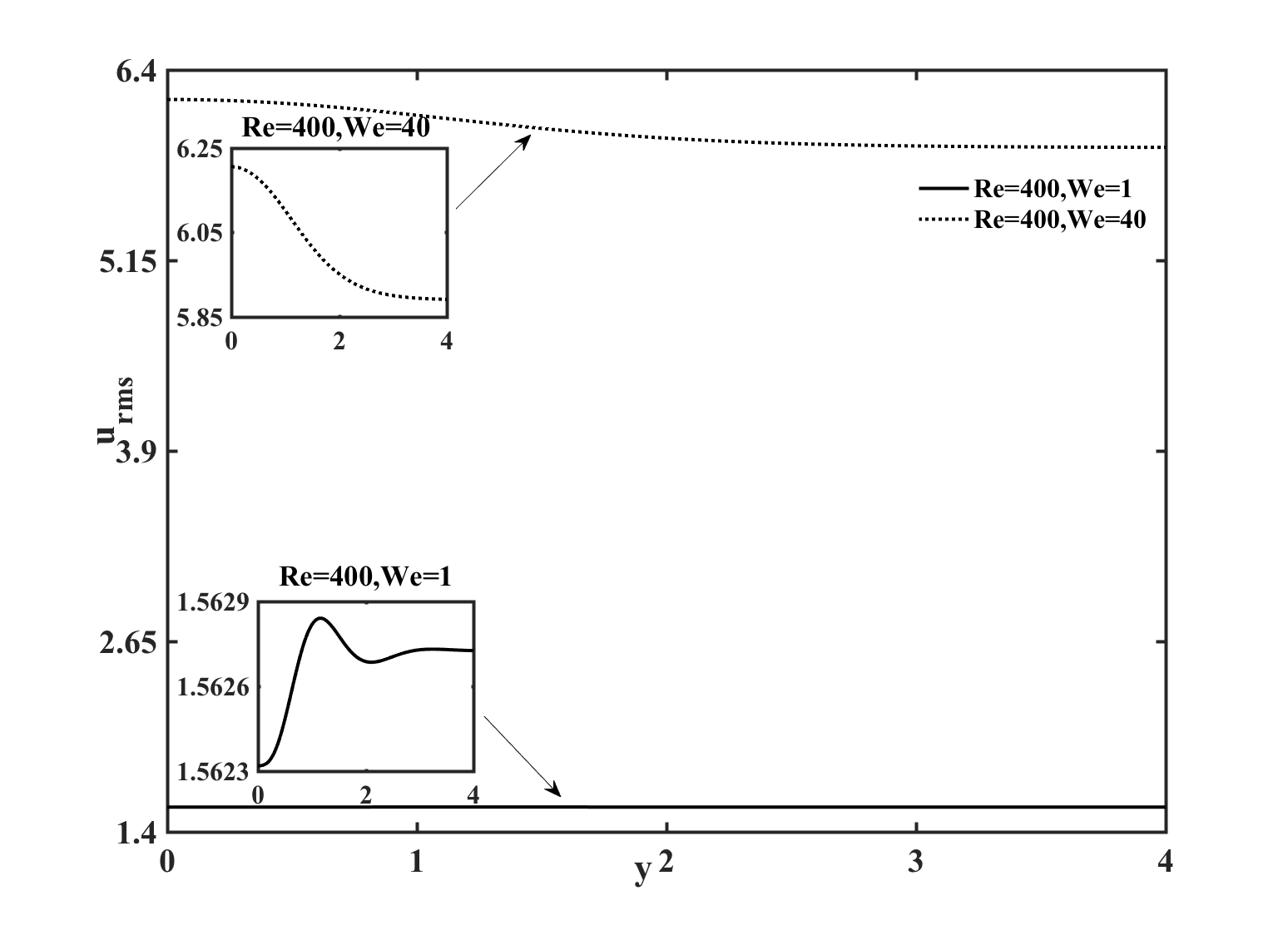}
  \subcaption{ }
  \label{fig:Fig4a}
\end{subfigure}
\hspace*{\fill}%
\begin{subfigure}{.5\textwidth}
  \centering
  \includegraphics[width=.9\linewidth, height=0.75\linewidth]{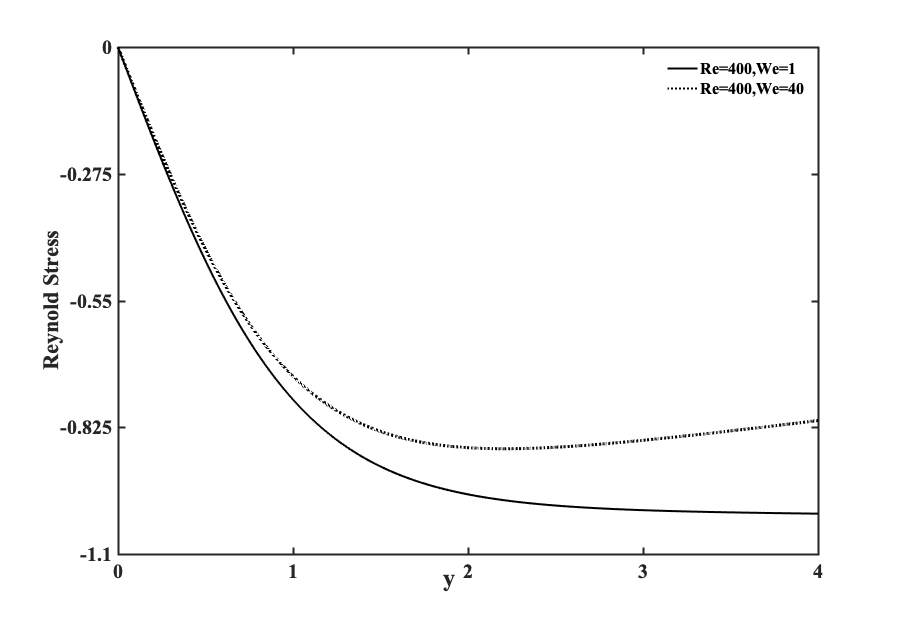}
  \subcaption{ }
  \label{fig:Fig4b}
\end{subfigure}
\caption{(a) Streamwise root mean velocity fluctuations, ${\displaystyle u_{\text{rms}}(0,y) = \sqrt{\frac{1}{T}\int^T_0 u'(0,y,t) dt}}$ and (b) Residual Reynolds shear stress, ${\displaystyle = \frac{1}{T}\left(\int^T_0 (U(y)+u'(0,y,t))v'(0,y,t) dt\right) - U(y)}$ evaluated at a fixed  streamwise location, $x_0=0$ and at the peak of the maximum growth rate curves (refer Figure~\ref{fig:Fig1} caption). The inset in (a) shows the maximum variation in ${\displaystyle u_{\text{rms}}(x,y)}$ at $y \approx 1$, the edge of the mixing layer. $u'(x,y,t)$ and $v'(x,y,t))$ are the streamwise and transverse velocity fluctuations (equation~\eqref{eqn:disturbance}) at $Re=400, We=1$ (solid curve), $Re=400, We=40$ (dotted curve) and at fixed viscosity coefficient, $\nu=0.5$. $U(y)$ is the mean flow (equation~\eqref{eqn:Meanflow}). ${\displaystyle T = (\nicefrac{2\pi}{\omega^\text{max}_r})}$ is one time period and $\omega^\text{max}_r$ is the real part of the angular frequency at the peak of the maximum growth rate.}\label{fig:Figure4}
\end{figure}

An equivalent outcome can be drawn by examining the root mean square of streamwise velocity fluctuations, $u_{\text{rms}}(0,y)$ (Figure~\ref{fig:Fig4a}) and Residual Reynolds shear stress (Figure~\ref{fig:Fig4b}) at a fixed  streamwise location, $x_0=0$. A larger drag reduction rate (associated with higher $We$) begets a larger $u_{\text{rms}}$ (e.~g., $u_{\text{rms}} \approx 1.5, 6.0$ at $We=1.0, 40$ respectively) and a lower residual Reynolds stresses in an absolute sense. To summarize the viscoelastic free shear flows are less unstable in the presence of viscoelastic additives.

%
\subsection{Spatio-temporal stability analysis} \label{subsec:space-time}
\begin{figure}[htbp]
\begin{subfigure}{.5\textwidth}
  \centering
  \includegraphics[width=.8\linewidth, height=0.7\linewidth]{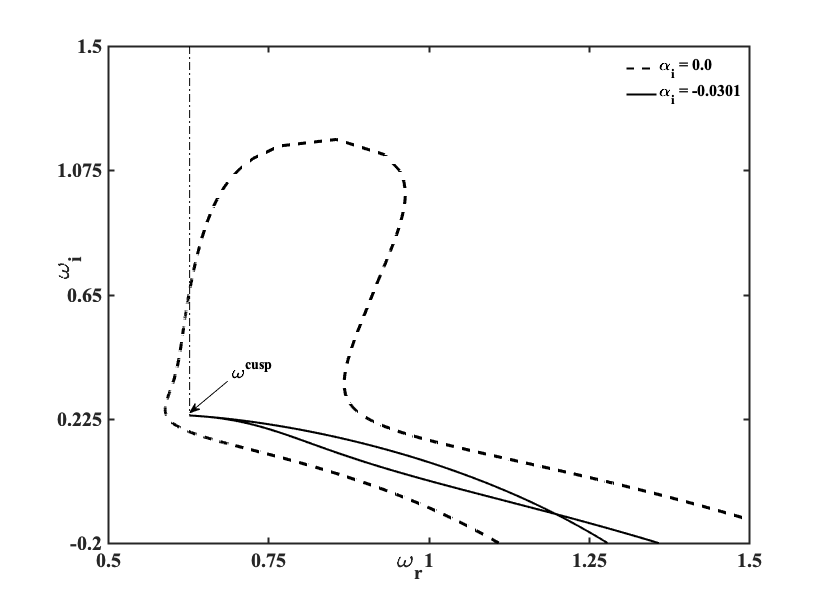}
  \subcaption{ }
  \label{fig:Fig5a}
\end{subfigure}
\hspace*{\fill}%
\begin{subfigure}{.5\textwidth}
  \centering
  \includegraphics[width=.8\linewidth, height=0.7\linewidth]{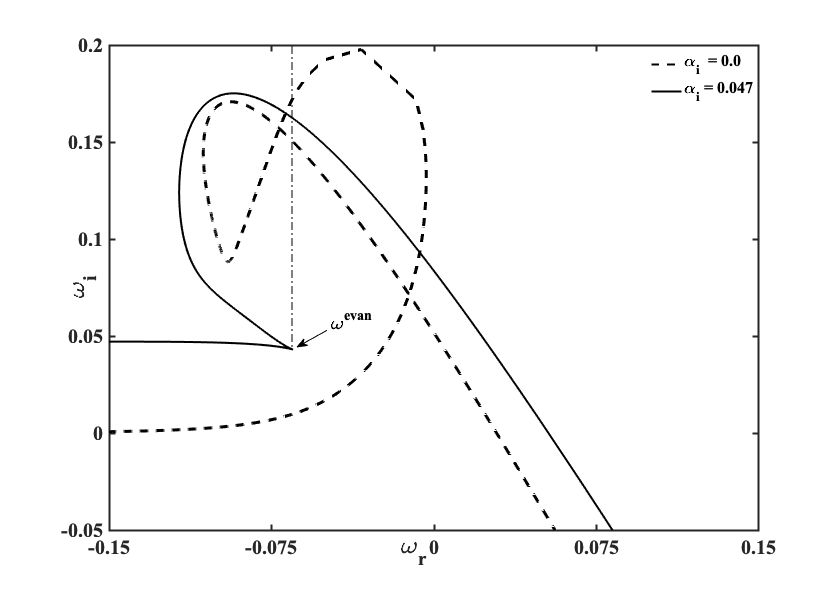}
  \subcaption{ }
  \label{fig:Fig5b}
\end{subfigure}
\caption{(a) Cusp point formation in $\omega-$plane for $Re=100, We=1, \nu=0.9$. The cusp point is formed at $\omega^{\text cusp}=0.6268 + 0.2362 i$ with the corresponding pinch point at $\alpha^{\text pinch}=-0.041 - 0.0301 i$. The ray drawn from the cusp point, parallel to the y-axis intersects the $\alpha_i=0$ curve once (odd number of times), indicating a genuine cusp point, (b) Evanescent mode formation for $Re=100, We=1, \nu=0.5$ at $\omega^{\text evan}=-0.0649 + 0.0430 i$ and with the corresponding pinch point at $\alpha^{\text pinch}=-0.038 + 0.047 i$. The ray drawn from this cusp point, parallel to the y-axis intersects the $\alpha_i=0$ curve twice (even number of times).}\label{fig:Figure5}
\end{figure}

\begin{figure}[htbp]
\begin{subfigure}{.5\textwidth}
  \centering
  \includegraphics[width=0.95\linewidth, height=0.7\linewidth]{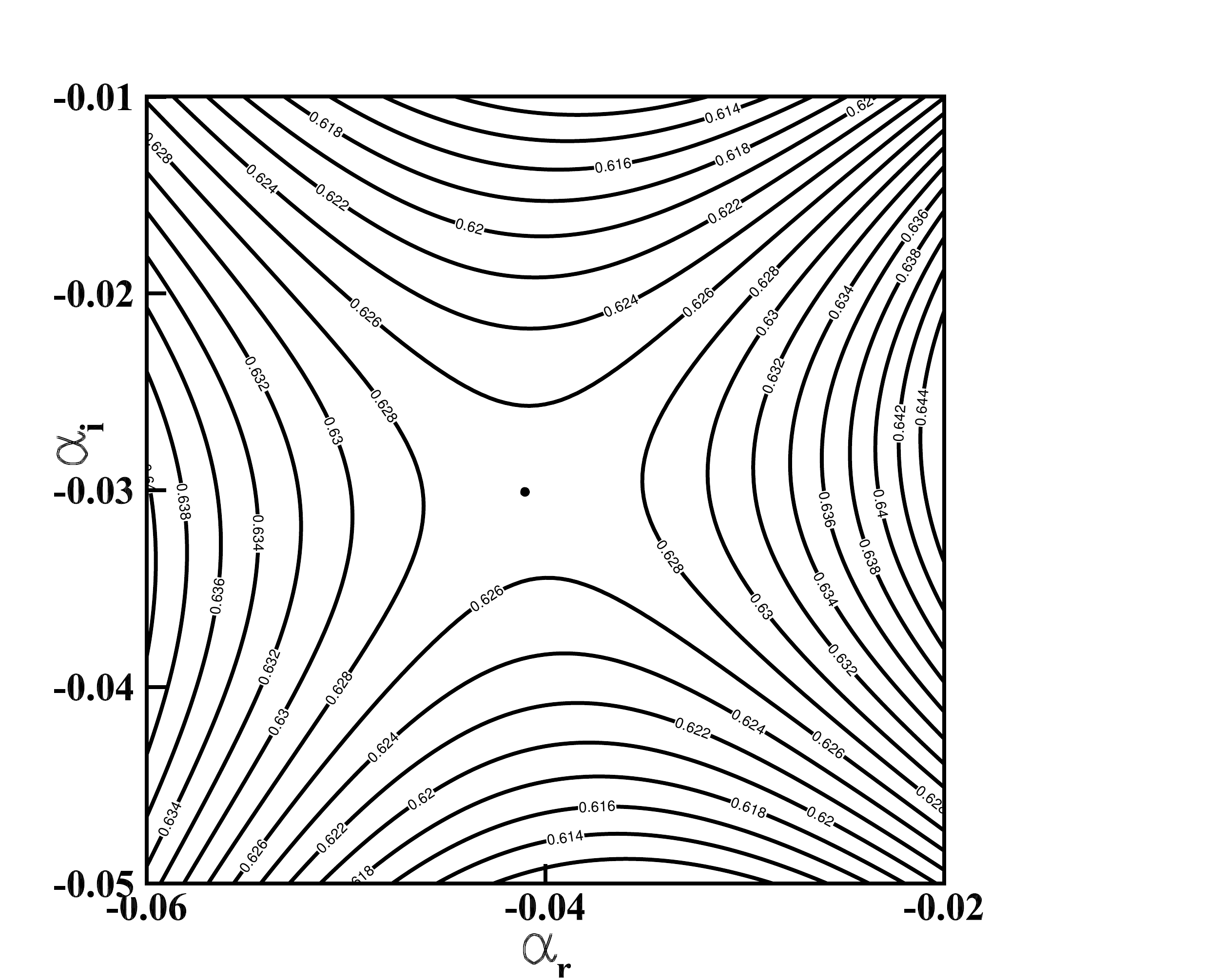}
  \subcaption{ }
  \label{fig:Fig6a}
\end{subfigure}
\hspace*{\fill}
\begin{subfigure}{.5\textwidth}
  \centering
  \includegraphics[width=.95\linewidth, height=0.7\linewidth]{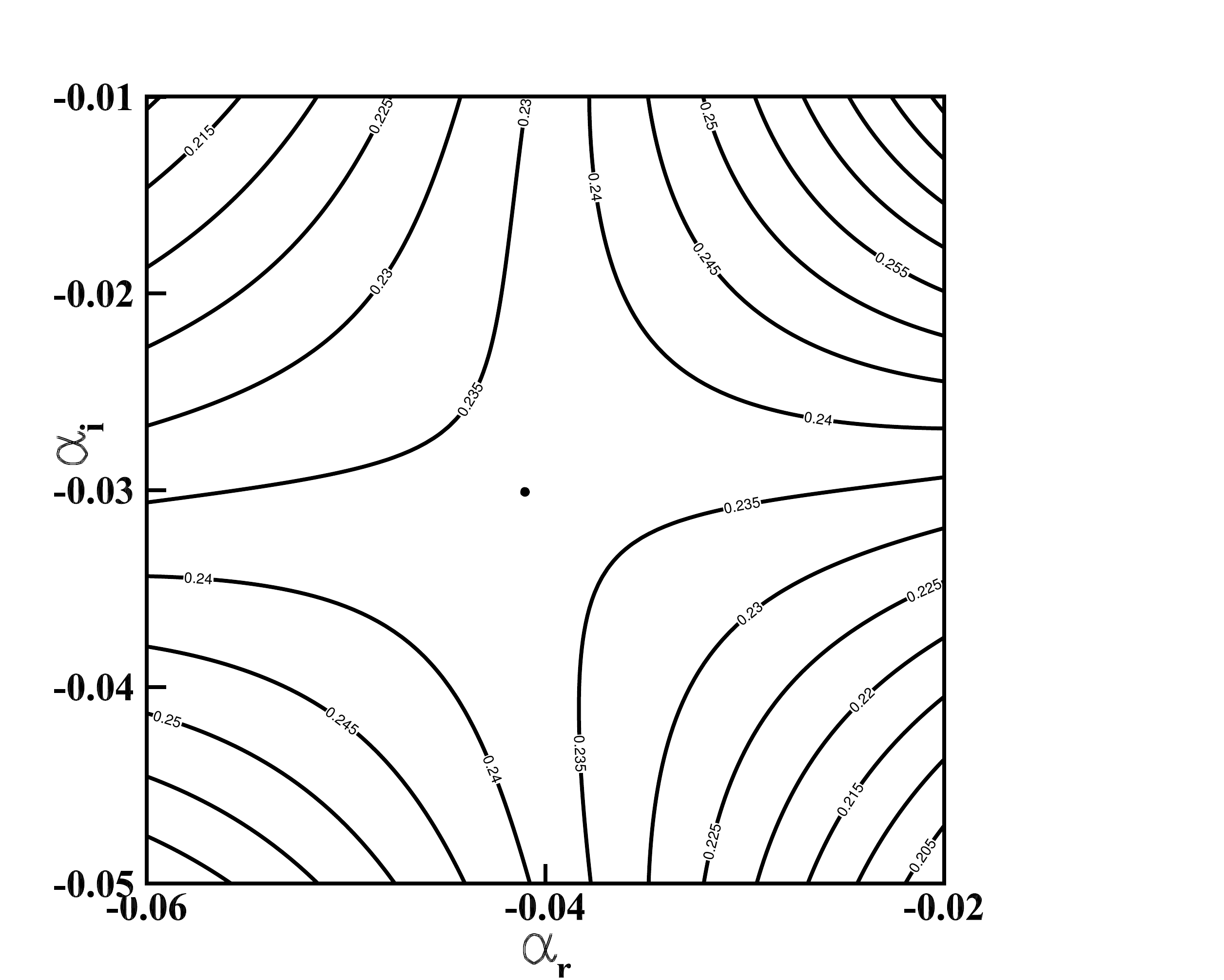}
  \subcaption{ }
  \label{fig:Fig6b}
\end{subfigure}
\caption{ 
The pinch point, $\alpha^{\text pinch}=-0.041 - 0.0301 i$, in the $\alpha-$plane corresponding to the cusp point as outlined in Figure~\ref{fig:Fig5a} for $Re=100, We=1, \nu=0.9$, which are demonstrated by drawing the iso-contours of (a) $\omega_r$ and (b) $\omega_i$.}\label{fig:Figure6}
\end{figure}

The spatio-temporal analysis, where both $\omega$ and $\alpha$ are designated as complex numbers, is a step which is necessary to reveal the stability phase diagram in the flow-material parameter space (i. e., in the ($Re, We, \nu$)-space, Figure~\ref{fig:Figure8}). In the onset, we demonstrate an example of a cusp point in the $\omega$-plane (Figure~\ref{fig:Fig5a}) and the saddle point in the $\alpha$-plane (Figure~\ref{fig:Figure6}) which correspond to each other via the local angle-doubling map as indicated by Kupfer~\cite{Kupfer1987} (details in \S~\ref{subsec:modelB}). To this end, a particular set of parameters are chosen for which the flow is absolutely unstable: $Re=100, We=1, \nu=0.9$. First, the image of the dispersion relation at $\alpha_i=0$ is obtained (Figure~\ref{fig:Fig5a}). This curve is the root of the DRP relation (equation~\eqref{eqn:OSE_DRP}) with the maximum positive imaginary component of the frequency, $\omega_i^{\text{max}}$, as outlined in \S \ref{subsec:time}. As we decrease the imaginary component of the wavenumber (for downstream mode in this example), the corresponding image in the $\omega$-plane is observed. When $\alpha_i$ is sufficiently negative (e.~g., $\alpha_i=-0.0301$), we detect the appearance of the cusp point at $\omega^{\text{cusp}}=0.6268+0.2362{\it i}$. The pinch point in the $\alpha$-plane (which is also the saddle point) corresponding to the cusp point in the $\omega$-plane arises at $\alpha^{\text{pinch}}=-0.041-0.0301{\it i}$ (Figure~\ref{fig:Figure6}). The pinch point is obtained by drawing the isocontours of $\omega_r$ (Figure~\ref{fig:Fig6a}) and $\omega_i$ (Figure~\ref{fig:Fig6b}) in the $\alpha$-plane. Consequently, by drawing a ray parallel to the $\omega_i$ axis (Figure~\ref{fig:Fig5a}), we note that this ray intersects $\alpha_i=0$ curve only once (i.~e., odd number of times), implying that the cusp point is genuine. Since the imaginary component of this cusp point is positive ($\omega_i^{\text{cusp}}=0.2362$), the free shear viscoelastic fluid flow at $Re=100, We=1, \nu=0.9$ is absolutely unstable.

Evanescent modes are also encountered in our analysis and to illustrate their existence we consider another example of the cusp point formation at $Re=100,\,\,We=1\,\,\,\nu=0.5$ (Figure~\ref{fig:Fig5b}). In this case, a ray emerging from the cusp point and parallel to the $\omega_i$ axis, intersects the $\alpha_i=0$ curve twice (i.~e., even number of times), thereby establishing that this cusp point corresponds to an evanescent mode. Evanescent modes (also known as the direct resonance mode~\cite{Koch1986}) arises if the two coalescing modes (given by the solution to equation~\eqref{eqn:DRP}) originate from waves propagating in the same direction (detailed discussion in \S \ref{subsec:modelB}). Linear stability theory predicts that these disturbances decay in the asymptotic limit of long time, but in the short term an algebraic growth associated with such a mode may be decisive to cause the transition to turbulence in receptivity studies~\cite{Lingwood1997}.

\begin{figure}[htbp]
\centering
\includegraphics[scale=0.15]{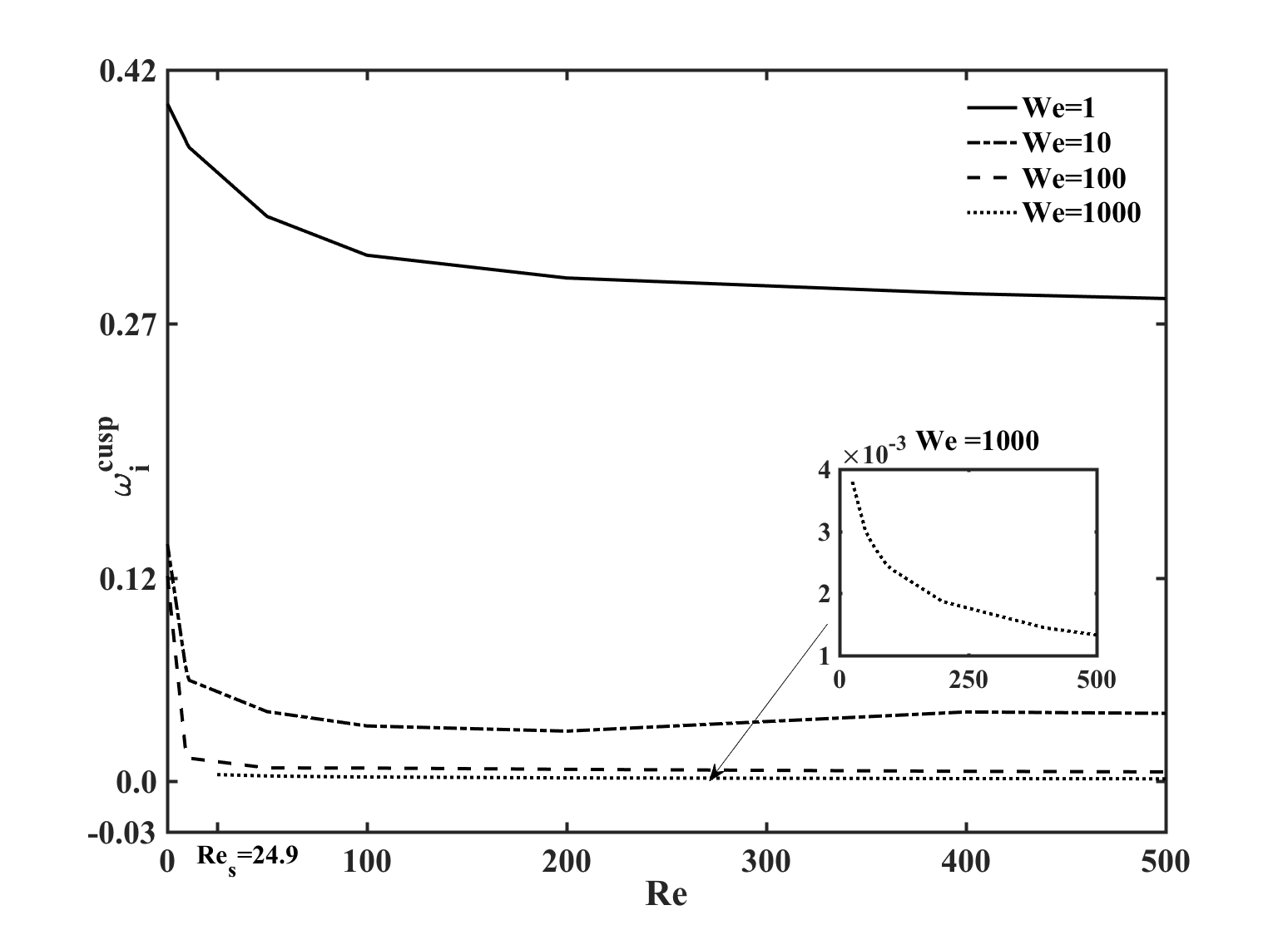}
  \vspace*{\fill}%
\caption{The cusp points, $\omega^{\text cusp}_i$ versus the Reynolds number in regions of convective and absolute instabilties calculated at the viscosity coefficient $\nu=0.35$.}\label{fig:Fig7}
\end{figure}

\begin{figure}[htbp]
\begin{subfigure}{.5\textwidth}
  \centering
  \includegraphics[width=0.95\linewidth, height=0.7\linewidth]{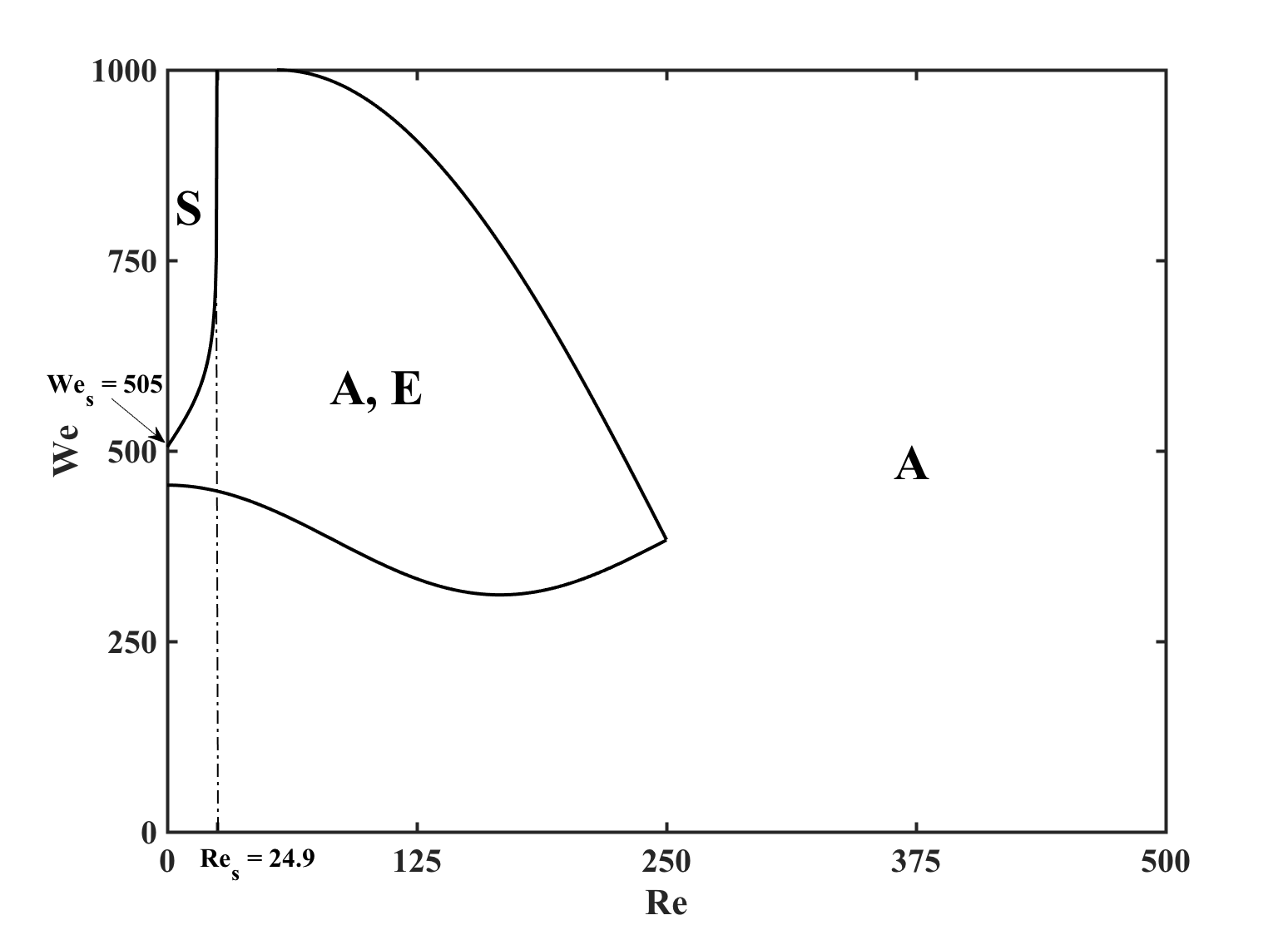}
  \subcaption{ }
  \label{fig:Fig8a}
\end{subfigure}
\hspace*{\fill}
\begin{subfigure}{.5\textwidth}
  \centering
  \includegraphics[width=.95\linewidth, height=0.7\linewidth]{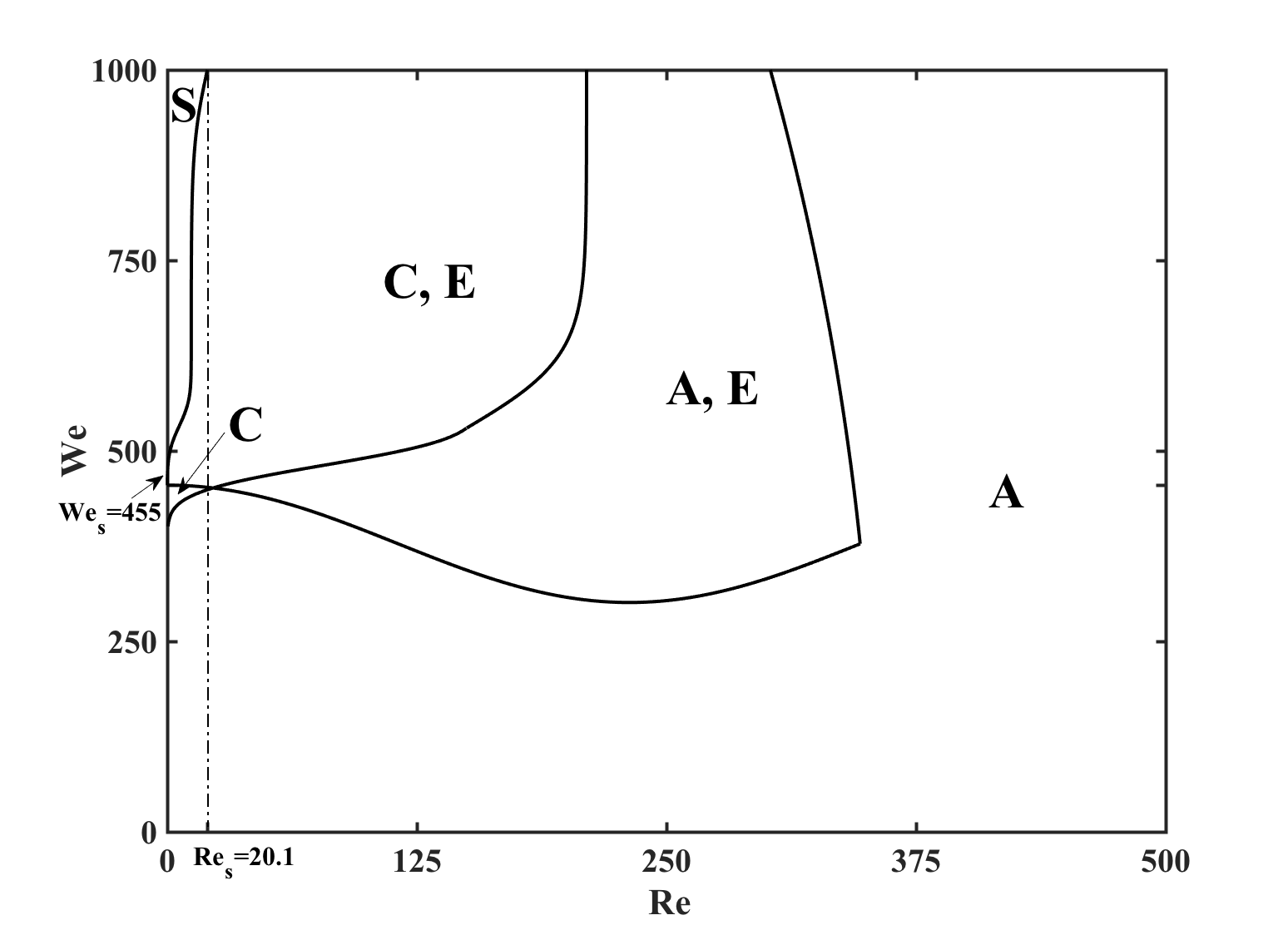}
  \subcaption{ }
  \label{fig:Fig8b}
\end{subfigure}
\caption{Viscoelastic free shear layer stability phase-diagram at (a) $\nu=0.35$ and (b) $\nu=0.20$, in the Re-We parametric space. The regions {\bf S, C, A} are denoted by temporally stable, convectively unstable and absolutely unstable regions, respectively. The domains outlined by {\bf A, E} and {\bf C, E} are those where both the unstable and evanescent modes (denoted by {\bf E}) are found.}\label{fig:Figure8}
\end{figure}

In order to determine the range of $Re$ (for fixed $\nu$, and different $We$) for which the flow regimes are absolutely unstable, convectively unstable or temporally stable, we plot the absolute growth rate in Figure~\ref{fig:Fig7} (i.~e., $\omega_i^{\text cusp}$, which is the growth rate at the cusp point obtained by deforming the Fourier contour in the $\alpha$-plane starting from the peak of the temporal growth rates, as explained in the above two examples). We define $Re_s$ ($We_s$) as the maximum critical Reynolds number (minimum critical Weissenberg number) for the flow to be temporally stable. Notice that the magnitude of the absolute growth rate (wherever it exists) predominantly decreases with increasing $Re$ and with increasing $We$. Although the combined effect of these parameters on the flow stability are often reported in terms of the elasticity number ($E_{\it l} = We/Re$, which corresponds to the ratio of the time scale for the elastic stress evolution to the time scale for diffusion of momentum~\cite{Azaiez1994}), we elaborate the impact of these flow-material parameters via the stability phase-diagram, discussed next.

We depict the boundaries of the temporally stable regions ({\bf S}), convective instabilities ({\bf C}), evanescent modes ({\bf E}) and absolute instabilities ({\bf A}) within a selected range of flow-material parameter space, i.~e., $Re \in [0.1, 500]$, $We \in [1, 1000]$ and $\nu=0.35$ (Figure~\ref{fig:Fig8a}), $\nu=0.20$ (Figure~\ref{fig:Fig8b}). The boundaries of temporally stable regions are estimated to reside within the range $We \in [We_s=505, 1000]$, $Re \in [0, Re_s=24.9]$ at $\nu=0.35$ (refer Figure~\ref{fig:Fig8a}) and $We \in [We_s=455, 1000]$, $Re \in [0, Re_s=20.1]$ at $\nu=0.20$ (refer Figure~\ref{fig:Fig8b}), respectively. The linear stability phase diagram for the free shear flow systems dominated by viscous stresses (i.~e., the case in which $\nu > 0.35$), have revealed only the absolutely unstable and the evanescent modes within an identical range of the flow parameters. Hence, the discussion of those phase diagrams are omitted in this article. Two conclusions can be deduced from these phase diagrams. First, since the boundary of the absolutely unstable region, ({\bf A, E}), is not parallel to the y-axis, the authors surmise that the flow will be absolutely unstable for sufficiently high $Re$ and irrespective of $We$. This assertion is supported by previous numerical studies of inviscid viscoelastic mixing layer in the limit of high $Re, We$ such that the elasticity number is held fixed and finite~\cite{Azaiez1994}. Second, unlike Newtonian fluids, viscoelastic liquids are either (absolutely/convectively) unstable for all $Re$ (e.~g., consider the region $We < 455$ in Figure~\ref{fig:Fig8a} and~\ref{fig:Fig8b}) or the transition to instability occurs at very low $Re$ (e.~g., $Re_s=24.9, 20.1$ at $\nu=0.35, 0.20$ and $We=1000$, respectively). Further in the latter case, the transition is direct from temporally stable region to absolutely unstable region for elastic stress dominated fluids (i.~e., the case of $\nu=0.35$, Figure~\ref{fig:Fig8a}) while this transition is more gradual, ${\bf S}\rightarrow{\bf C}\rightarrow{\bf A}$ for viscous stress dominated fluids (i.~e., the case of $\nu=0.20$, Figure~\ref{fig:Fig8b}). Although absolute instability is far more vicious than convective instability as the occurrence of absolute instability pervades the entire flow domain, convective instability can also cause transition to turbulence if sufficient spatial distance is given for the disturbance to grow~\cite{Huerre1990}. The second conclusion can be attributed to the fact that viscoelasticity dramatically exacerbates free surface flow instabilities~\cite{Samanta2013}. This destabilization arises from a combination of large tangential tension and the curvature along the free surface leading to an unstable stress gradient.
%

We recapitulate our discussion by indicating some experimental findings which corroborate our numerical results, starting from the polymer drag reduction predictions in the early 1980s by Hibbert who used Laser-Doppler anemometry to notice the tendency for the enhancement of the vorticity structures and the diminution of small scale turbulence in free shear flow of polymeric liquids~\cite{Hibberd1982}. The flow visualization experiments of Riediger on surfactant blended fluids showed a longer lifetime of the typical structures with the mixing layer, i.~e. roll-up and pairing, as well as a tendency for the suppression of the mixing layer growth~\cite{Riediger1989}. More recently conducted experiments by Rahgozar et al.~\cite{Rahgozar2017} have reported an order of magnitude reduction in the natural (unforced) transition $Re$ of boundary-free, uniformly sheared, dilute polymer solutions; a phenomena rebaptized as `early turbulence'~\cite{Samanta2013}. Subsequently, this new transitional pathway to the elastoinertial turbulent state have been experimentally demonstrated to exist over a much wider range of flow-material parameters~\cite{Varshney2018}.

\section{Conclusions} \label{sec:conclusion}
This investigation addresses the linear, temporal and spatio-temporal analyses of free shear flows of dilute polymeric liquids with anti-symmetric centerline conditions for low to moderate Reynolds and Weissenberg numbers. \S \ref{sec:model} presented the free shear viscoelastic flow as well as the elements of the linear stability analysis via the solution of the Orr Sommerfeld equation, including the illustration of the Briggs contour integral method to determine the existence of absolutely unstable, convectively unstable and evanescent modes. \S \ref{sec:CMM} demonstrated the steps of the Compounded Matrix Method, utilized to numerical solve the resultant system of stiff differential equations. The numerical outcome, in a selected range of Reynolds and Weissenberg number, highlight that with increasing Weissenberg number the peak of the maximal growth rate is reduced, the region of temporal instability is gradually concentrated near zero wavenumber, the vorticity structures of the disturbance become larger and more widely spaced and the residual Reynolds stresses are lowered in absolute sense, all these observations indicating a mechanism of elastic stabilization in non-Newtonian free shear flows of dilute solutions. The spatio-temporal analysis reveal that either the flow is unstable for all Reynolds number or the transition to instability transpires at very low Reynolds number stipulating the significance of the unraveling of the polymer chains at the initiation of the free surface instabilities. Although the Oldroyd-B model describes well the behavior of polymeric liquids composed of a low concentration of high molecular weight polymer in a very viscous Newtonian solvent at moderate shear rates, more realistic description of polymeric solution and melts, including the effect of shear thinning, non-zero normal stress coefficient and stress overshoot in transient flows is missing and hence deserves a full numerical exploration in the near future, using a constitutive relation which incorporates these above mentioned features.

\paragraph{Acknowledgements:} Authors thank Prof.~Tapan Sengupta (Dept. Aerospace Engg, IIT Kanpur) for introducing the Compound Matrix Method. SS acknowledges the financial support of the grant DST ECR/2017/000632. DB acknowledges the financial support of her fellowship CSIR 09/1117(0004)/2017-EMR-I which helped her run the numerical simulations.


\end{document}